\documentclass[twocolumn,showpacs,preprintnumbers,amsmath,amssymb]{revtex4-1}
\usepackage[whole]{bxcjkjatype} 
\usepackage[unicode]{hyperref} 
\usepackage{framed}
\usepackage{graphicx}
\usepackage{dcolumn}
\usepackage{bm}
\usepackage[compact]{titlesec}

\renewcommand{\arraystretch}{1.6}       
\usepackage{xcolor}

\newcommand{\half}{{{\textstyle\frac{1}{2}}}}
\newcommand{\quarter}{{{\textstyle\frac{1}{4}}}}
\newcommand{\be}{\begin{equation}}
\newcommand{\ee}{\end{equation} }
\newcommand{\beqa}{\begin{eqnarray} }
\newcommand{\eeqa}{\end{eqnarray} }
\newcommand{\ba}{\begin{array}}
\newcommand{\ea}{\end{array}}
\newcommand{\bpm}{\begin{pmatrix}}
\newcommand{\epm}{\end{pmatrix}}

\newcommand{\ccLR}{\mathbf{c}_{{\scriptscriptstyle{\bf{L/R}}}}}
\newcommand{\ccL}{\mathbf{c}_{{\scriptscriptstyle{\bf{L}}}}}
\newcommand{\ccR}{\mathbf{c}_{{\scriptscriptstyle{\bf{R}}}}}
\newcommand{\QL}{Q_{{\scriptscriptstyle{\bf{L}}}}}
\newcommand{\QR}{Q_{{\scriptscriptstyle{\bf{R}}}}}
\newcommand{\QB}{Q_{{\scriptscriptstyle{\bf{B}}}}}
\newcommand{\deltaB}{\delta_{{\scriptscriptstyle{\bf{B}}}}}

\newcommand{\rmd}{{\rm d}}

\newcommand{\rmD}{{\rm D}}

\newcommand\hcL{{\hat{\cal L}}}

\newcommand{\ODD}{\mathbf{O}(D,D)}

\newcommand\To{T_{\scriptscriptstyle{{(0)}}}}
\newcommand\So{S_{\scriptscriptstyle{{(0)}}}}

\newcommand\cH{{\cal H}}

\newcommand\cJ{{\cal J}}

\newcommand\cL{{\cal L}}

\newcommand\cO{{\cal O}}

\def\bri{\bar{\imath}}
\def\brj{\bar{\jmath}}

\def\tx{\tilde{x}}

\def\tB{\tilde{B}}
\def\tC{\tilde{C}}

\newcommand\tkappa{{\tilde{\kappa}}}

\def\na{\nabla}

\def\brbeta{\bar{\beta}}
\def\brgamma{\bar{\gamma}}

\def\brk{{\bar{k}}}
\def\brl{{\bar{l}}}

\def\brn{{\bar{n}}}

\def\brX{\bar{X}}
\def\brY{\bar{Y}}

\def\alphap{\alpha^{\prime}}

\newcommand\hC{\hat{C}}

\begin{document}
\preprint{YITP-20-037}
\title{String Theory and non-Riemannian Geometry}
\author{Jeong-Hyuck Park (朴\,廷\,爀)${}^{1,2}$}
\email{park@sogang.ac.kr}\email{On sabbatical leave of absence from $1$.}
\author{Shigeki Sugimoto (杉本~茂樹)${}^{2}$}
\email{sugimoto@yukawa.kyoto-u.ac.jp }

\affiliation{${}^{1}$Department of Physics, Sogang University, 35 Baekbeom-ro, Mapo-gu, Seoul 04107,  Korea\\
${}^{2}$Center for Gravitational Physics,  
Yukawa Institute for Theoretical Physics, Kyoto University, Kitashirakawa Oiwakecho, Sakyo-ku, Kyoto 606-8502, Japan}

\begin{abstract}
\centering\begin{minipage}{\dimexpr\paperwidth-6.2cm}
\noindent 
The $\mathbf{O}(D,D)$  covariant generalized metric, postulated  as a truly fundamental  variable, can describe  novel geometries  where the notion of Riemannian metric ceases  to exist.  Here we quantize a closed string upon  such backgrounds and identify flat, anomaly-free,   non-Riemannian string vacua in the familiar critical dimension, $D{=26}$ (or $D{=10}$). Remarkably,  the whole BRST closed string spectrum is restricted to just  one level  with no tachyon, and matches the linearized  equations of motion of Double Field Theory. Taken as an internal space,  our non-Riemannian vacua may open up  novel avenues alternative to  traditional  string compactification.
\end{minipage}
\end{abstract}

                             
\maketitle

\section*{Motivation: absence of tachyon kinetic term in DFT}\vspace{-3pt}
\noindent
Ever since the adoption of Riemannian geometry into the  formulation of  General Relativity,   the metric, $g_{\mu\nu}$, has been privileged  to be the fundamental  variable that  provides a concrete mathematical tool to  address the notion   of  \textit{spacetime}.  In particular, the `flat' spacetime where the  gravitational effect is negligible is simply given  by  a   constant  metric of  Minkowskian signature. Needless to say, the Standard Model upon this background is arguably  the best tested theory.   

The flat Minkowskian spacetime  is famously   known to be  unstable   in  bosonic string theory, as  both  the open and closed  string spectra  contain a negative mass-squared tachyon.  Open string tachyon means the instability of a D-brane. Its  tachyon  potential     has a local minimum which corresponds to   a closed string vacuum without any D-brane~\cite{Sen:1999mh,Sen:1999nx,Gerasimov:2000zp,Kutasov:2000aq,Schnabl:2005gv}. Yet,  for the closed string tachyon,  little is known  except  the effective description,\vspace{-3pt}
\be
\ba{l}
\!\displaystyle{\int}\rmd^{D}x~\sqrt{-g}e^{-2\phi}
\Big[R+4\partial_{\mu}\phi\partial^{\mu}\phi-\textstyle{\frac{1}{12}}H_{\lambda\mu\nu}H^{\lambda\mu\nu}\\
\,\qquad\,\,\,\,\,\,-\textstyle{\frac{2(D-26)}{3\alphap}}
-\partial_{\mu}T\partial^{\mu}T+\textstyle{\frac{4\,}{\alphap}}T^{2}+\cO(T^{3})\Big]\,.
\ea
\label{Reff}
\ee
\indent In this Letter, we construct  a novel bosonic closed string  theory with  a finite spectrum free of  tachyon, by going beyond the Riemannian paradigm. Our working hypothesis,   motivated by   T-duality,  is  to view an $\ODD$ invariant  metric, $\cJ_{MN}$, and an  $\ODD$ covariant generalized metric, $\cH_{MN}$, as  fundamental  entities (instead of  $g_{\mu\nu}$).  The former is put in an off-block diagonal form, $\cJ_{MN}={\tiny{\left(\ba{cc}\mathbf{0}&\mathbf{1}\\\mathbf{1}&\mathbf{0}\ea\right)}}$, and, with its inverse, lowers and raises the  $\ODD$  indices, capital Roman $M,N=1,2,\cdots,{2D}$. The latter then satisfies     twofold \textit{defining properties}:
\be
\ba{ll}
\cH_{MN}=\cH_{NM}\,,\qquad&\qquad
\cH_{M}{}^{K}\cH_{N}{}^{L}\cJ_{KL}=\cJ_{MN}\,.
\ea
\label{defH}
\ee
A famous parametrization  reads  \cite{Giveon:1988tt}
\be
\cH_{MN}=\left(\ba{cc}g^{\mu\nu}~&~-g^{\mu\sigma}B_{\sigma\lambda}~~\\
B_{\kappa\rho}g^{\rho\nu}~&~~g_{\kappa\lambda}-B_{\kappa\rho}g^{\rho\sigma}B_{\sigma\lambda}\ea\right)\,,
\label{Riemann}
\ee
which contains   $g_{\mu\nu}$ and a  skew-symmetric $B$-field, corresponding to a well-known     coset, ${\frac{\ODD}{\mathbf{O}(1,D{-1})\times{\mathbf{O}(D{-1},1)}}}$ for   Minkowskian signature.  Over the last three decades, this has been a cornerstone for the developments of    $\ODD$ symmetry manifest  formulations  of      worldsheet string theory~\cite{Duff:1989tf,Tseytlin:1990nb,Tseytlin:1990va,Rocek:1991ps,Giveon:1991jj,Hull:2004in,Hull:2006qs,Hull:2006va} and    also  target spacetime effective descriptions~\cite{Siegel:1993xq,Siegel:1993th,Hull:2009mi,Hull:2009zb,Hohm:2010jy,Hohm:2010pp}. They go under the name doubled string or  Double Field Theory (DFT), as the spacetime coordinates are formally  doubled, $x^{\mu}\rightarrow x^{M}=(\tx_{\mu},x^{\nu})$.

With (\ref{Riemann}), the closed string effective action~(\ref{Reff})  can be reformulated as  a DFT coupled to the tachyon,
\be
\!{\displaystyle{\int}}\!e^{-2d}
\Big[\So-\textstyle{\frac{2(D-26)}{3\alphap}}
-\cH^{MN}\partial_{M}T\partial_{N}T+\textstyle{\frac{4\,}{\alphap}}T^{2}+\cO(T^{3})\Big]\,.
\label{DFTeff}
\ee
Here $d$ is the $\ODD$ singlet  DFT-dilaton related to the  conventional dilaton through $e^{-2d}=\sqrt{-g}e^{-2\phi}$, and $\So$ denotes the  scalar curvature in  DFT which can be spelled out explicitly in terms of $d$, $\cH_{MN}$, and $\cJ_{MN}$~\cite{Hohm:2010pp}:
\[
\ba{l}
\So\!=\textstyle{\frac{1}{8}}\cH^{MN}\partial_{M}\cH_{KL}\partial_{N}\cH^{KL}+\half\cH^{MN}
\partial^{K}\cH_{ML}\partial^{L}\cH_{NK}\\
\qquad-\partial_{M}\partial_{N}\cH^{MN}
+4\partial_{M}(\cH^{MN}\partial_{N}d)-4\cH^{MN}\partial_{M}d\partial_{N}d\,.
\ea
\]
The so-called section condition should be imposed on the doubled coordinates, $\partial_{M}=(\tilde{\partial}^{\mu},\partial_{\nu})$,
\be
\partial_{M}\partial^{M}=\partial_{\mu}\tilde{\partial}^{\mu}+\tilde{\partial}^{\mu}\partial_{\mu}\equiv0\,,
\label{seccon}
\ee
such that   all quantities  have  $D$-dimensional  halved  dependence. After   solving the section condition by  letting $\tilde{\partial}^{\mu}{\equiv0}$ and assuming the Riemannian parametrization~(\ref{Riemann}), the DFT action~(\ref{DFTeff}) reduces to (\ref{Reff}). 

Crucially,  (\ref{Riemann}) is not the most general  solution to the defining relations~(\ref{defH}). It only becomes so, if the upper left ${D{\times D}}$ block, \textit{i.e.}$\cH^{\mu\nu}$,  is invertible. DFT  works perfectly  fine with any generalized metric that satisfies  (\ref{defH}).  In particular,  the ${D{\times D}}$ block can be degenerate, hence \textit{non-Riemannian} by nature~\cite{Lee:2013hma,Ko:2015rha,Morand:2017fnv,Cho:2018alk,Berman:2019izh,Cho:2019ofr,Blair:2020ops}  (see also \cite{Malek:2013sp,Park:2016sbw,Fernandez-Melgarejo:2017oyu,Sakatani:2019jgu,Blair:2019qwi,Berman:2020tqn} for  supersymmetric or exceptional examples). 
The most general  parametrizations of a generalized metric have  been classified by two non-negative integers, $(n,\brn)$, which set  $\mathrm{dim}(\mathrm{ker}\,\cH^{\mu\nu})={n+\brn}$, and  render string chiral and anti-chiral over $n$ and $\brn$ directions respectively~\cite{Morand:2017fnv}, see also (\ref{recast}) later. The Riemannian geometry of (\ref{Riemann})  is of $(0,0)$ type,  and non-relativistic/ultra-relativistic     strings~\cite{Gomis:2000bd,Danielsson:2000gi,Andringa:2012uz,Harmark:2017rpg,Bergshoeff:2014jla,Duval:2014uoa,Bekaert:2015xua} belong to $(1,1)$ or other  non-Riemannian  types~\cite{Ko:2015rha,Morand:2017fnv,Berman:2019izh,Blair:2019qwi}.

Postulating $\{\cJ_{MN},\cH_{MN},d\}$  as the only geometric quantities,  one can uniquely identify  a  covariant derivative, $\na_{M}$~\cite{Jeon:2010rw,Jeon:2011cn}, and then construct    the scalar curvature, $\So$,   
as well as  `Einstein' tensor, $G_{MN}$, which is off-shell conserved, ${\na_{M}G^{MN}=0}$~\cite{Park:2015bza}. This is all analogous  to General Relativity, though there seems no four-index  Riemann tensor~\cite{Hohm:2011si}. Using them, one can concisely   express all   the equations of motion of the DFT action~(\ref{DFTeff}):
\be
\ba{ll}
\cH^{MN}\na_{M}\na_{N}T+\textstyle{\frac{4\,}{\alphap}}T+\cO(T^{2})=0\,,\quad&\quad G_{MN}=T_{MN}\,.
\ea
\label{EOM}
\ee
The former is the tachyonic equation of motion and the   latter is the     `Einstein  equations' in DFT~\cite{Angus:2018mep,Park:2019hbc}. It  
unifies the equations of motion of $\cH_{MN}$ and $d$ into a single formula,    equating     the Einstein tensor  with a generalised  stress-energy tensor.  For  the tachyon field it reads
\[
T_{MN}=(\cJ{+\cH})_{[M}{}^{K}(\cJ{-\cH})_{N]}{}^{L}\partial_{K}T\partial_{L}T-\half\cJ_{MN}\To\,,
\]
where $\To=-\frac{1}{D}T_{M}{}^{M}$ is the $\ODD$ singlet trace part,
\[
\To=\textstyle{\small{\cH^{MN}\partial_{M}T\partial_{N}T-\textstyle{\frac{4\,}{\alphap}}T^{2}-\cO(T^{3})+\textstyle{\frac{2(D-26)}{3\alphap}}}}\,.
\]
In particular, the equation of motion of the DFT-dilaton, or  the trace of the Einstein equations,  implies ${\So=\To}$. Thus, in ${D=26}$, if the tachyon potential admits a global minimum away from ${T=0}$, we have ${\To<0}$   and hence the background cannot be flat, ${\So<0}$.  

While we refer the interested readers to  section~2 of \cite{Angus:2018mep} for  a  detailed  review of the above formalism, for now what suffices us  is that the Einstein curvature, $G_{MN}$, vanishes for constant $\cH_{MN}$ and  $d$.   Any flat background with vanishing tachyon, ${T=0}$, solves  all the equations of motion~(\ref{EOM})  in the critical dimension, ${D=26}$. Surely this statement is also valid  for the Riemannian action~(\ref{Reff}). The novelty here is  that   the DFT action~(\ref{DFTeff}) allows  non-Riemannian geometries. With the choice of the section by  ${\tilde{\partial}^{\mu}\equiv 0}$,   the tachyon kinetic term reads $\cH^{\mu\nu}\partial_{\mu}T\partial_{\nu}T$ which obviously vanishes when  ${\cH^{\mu\nu}=0}$.  The vanishing   kinetic term then  may   eliminate the  instability of the static configurations: there is no dynamics for the tachyon to roll down  (at least classically at linear order).    
The absence of kinetic term was also discussed  for tachyon condensation in open string field theory~\cite{Rastelli:2000hv,Hata:2001rd}, while it is a generic feature of 
`Pregeometrical' or `Purely Cubic'  string field theories~\cite{Hata:1986vq,Horowitz:1986dta}. 

The generalised metric with  $\cH^{\mu\nu}{=0}$  is in a way `maximally' non-Riemannian, invalidating any notion of Riemannian metric, not to mention its signature. It corresponds to the $(n,\brn)$ type with ${n+\brn=D}$,   and  assumes the  most general form~\cite{Morand:2017fnv},
\be
\!\cH_{MN}=\!\!\left(\ba{cc}{0}&{Y_{i}^{\mu}X^{i}_{\lambda}-\brY_{\bri}^{\mu}\brX^{\bri}_{\lambda}}\\{
X^{i}_{\kappa}Y_{i}^{\nu}-\brX^{i}_{\kappa}\brY^{\nu}_{i}}&{
2X^{i}_{(\kappa}B_{\lambda)\rho}Y_{i}^{\rho}-2\brX^{\bri}_{(\kappa}B_{\lambda)\rho}\brY_{\bri}^{\rho}
}\ea\right),
\label{cHcJ}
\ee
where  $i=1,2,\cdots, n$ and $\bri=1,2,\cdots,\brn$. Viewing  $(X_{\mu}^{i},\brX_{\mu}^{\bri})$  as a $D\times D$  matrix, $(Y^{\nu}_{i},\brY^{\nu}_{\bri})$ is its inverse  satisfying $X_{\mu}^{i}Y^{\nu}_{i}+\brX_{\mu}^{\bri}\brY^{\nu}_{\bri}=\delta_{\mu}{}^{\nu}$. The underlying coset is ${\frac{\ODD}{\mathbf{O}(n,n)\times{\mathbf{O}(\brn,\brn)}}}$~\cite{Berman:2019izh} whose    dimension,   $4n\brn$, matches  the number of   infinitesimal fluctuation modes, \textit{i.e.~}moduli, around the generalized metric~(\ref{cHcJ})~\cite{Cho:2019ofr}. The types of $(D,0)$ or $(0,D)$ are worthy of note. They are uniquely given by $\cH_{MN}=\pm\cJ_{MN}$, and correspond to  the most   symmetric  vacua of DFT with no moduli~\cite{Cho:2018alk}.    Intriguingly then,    Riemannian spacetime may arise  in DFT after the spontaneous symmetry breaking of  $\ODD$,  which identifies   $g_{\mu\nu}$ and $B_{\mu\nu}$   as the  massless  Nambu--Goldstone bosons~\cite{Berman:2019izh} (\textit{c.f.~}\cite{Kirsch:2005st,Alexander:2016xuy}).

In the remainder of this Letter, we investigate the quantum consistency  of the non-Riemannian  geometries~(\ref{cHcJ}) as  for novel  string vacua. Through BRST quantization of the string, we show that  the type of $(n,\brn)=(13,13)$ with ${D=26}$ is  anomaly-free. Remarkably,  the string spectrum  is  \textit{finite} with no tachyon mode,   matches the  coset underlying (\ref{cHcJ}),  and agrees with the linearized    DFT  equations of motion, \textit{i.e.}~the vacuum Einstein  equations, ${G_{MN}=0}$. We shall  conclude with remarks on  extension to  type II superstring and   application as an alternative  to   string compactification on Riemannian manifolds.


\section*{BRST quantization of doubled-yet-gauged string\label{SECw}}
\noindent
The  doubled  string action we consider is \cite{Hull:2006va,Lee:2013hma},
\be
\ba{l}
S=\frac{1}{4\pi\alphap}\displaystyle{\int}\rmd^{2}\sigma~\cL\,,\\
\cL=-\half\sqrt{-h}h^{\alpha\beta}\rmD_{\alpha}x^{M}\rmD_{\beta}x^{N}\cH_{MN}
-\epsilon^{\alpha\beta}\rmD_{\alpha}x^{M}A_{\beta M}\,.
\ea
\label{stringaction}
\ee
${\rmD_{\alpha}}$ is a  covariant derivative with     
an auxiliary  potential that satisfies a section-condition-like constraint, 
\be
\ba{ll}
{\rmD_{\alpha} x^{M}:=\partial_{\alpha}x^{M}-A_{\alpha}^{M}}\,,\qquad&\quad
A_{\alpha}^{M}\partial_{M}=0\,.
\ea
\label{rmDz}
\ee
While the action is completely covariant under   desired symmetries like  $\ODD$ rotations, Weyl symmetry,  worldsheet as well as doubled target spacetime  diffeomorphisms, it also       concretely    realizes the idea  that the  doubled coordinates in DFT are actually gauged and each gauge orbit corresponds to a single physical point~\cite{Park:2013mpa}. The  relevant   `coordinate gauge symmetry' reads 
\be
\ba{lll}
\delta x^{M}=\Delta^{M}\,,\qquad&\quad\delta A_{\alpha}^{M}=\partial_{\alpha}\Delta^{M}\,,\qquad&\quad\Delta^{M}\partial_{M}=0\,,
\ea
\label{CGS}
\ee
which leaves $\rmD_{\alpha}x^{L}$, $\cH_{MN}$  invariant (${\Delta^{L}\partial_{L}\cH_{MN}=0}$), and enables us to  identify  the first term in the Lagrangian~(\ref{stringaction}) as  a  `proper area' in doubled geometry~\cite{Park:2017snt}.

With the choice of the section,  
$(\tilde{\partial}^{\mu},\partial_{\nu})\equiv(0,\partial_{\nu})$, 
which we henceforth assume throughout, the constraints  on the gauge potential~(\ref{rmDz})  and  parameter~(\ref{CGS}) are solved by $A_{\alpha}^{M}\equiv(A_{\alpha\mu},0)$ and
 $\Delta^{M}\equiv(\Delta_{\mu},0)$.  Clearly then, it is   the  tilde coordinates $\tx_{\mu}$ that  are to be   gauged: 
${\rmD_{\alpha} x^{M}=\big(\partial_{\alpha}\tilde{x}_{\mu}-A_{\alpha\mu}\,,\,\partial_{\alpha} x^{\nu}\big)}$.

Upon the Riemannian background~(\ref{Riemann}),  the   potential $A_{\alpha\mu}$  appears quadratically in  the action, $\textstyle{\frac{1}{4\pi\alpha^{\prime}}}\cL
=\textstyle{\frac{1}{2\pi\alpha^{\prime}}}
\cL^{\prime}$,
\be
\ba{l}
\cL^{\prime}=-\half\sqrt{-h}h^{\alpha\beta}\partial_{\alpha}x^{\mu}\partial_{\beta}x^{\nu}
g_{\mu\nu}
+\half\epsilon^{\alpha\beta}\partial_{\alpha}x^{\mu}
\partial_{\beta}x^{\nu}B_{\mu\nu}\\
+\half\epsilon^{\alpha\beta}
\partial_{\alpha}\tx_{\mu}\partial_{\beta}x^{\mu}-\quarter\sqrt{-h}h^{\alpha\beta}(A_{\alpha\mu}{-V_{\alpha\mu}})(A_{\beta\nu}{-V_{\beta\nu}})g^{\mu\nu}\,,
\ea
\label{Lpr}
\ee
where  
$V_{\alpha\mu}=\partial_{\alpha}\tx_{\mu}-B_{\mu\nu}\partial_{\alpha}x^{\nu}+\textstyle{\frac{1}{\sqrt{-h}}}\epsilon_{\alpha}{}^{\beta}g_{\mu\nu}\partial_{\beta}x^{\nu}$, and the worldsheet indices 
are raised/lowered with $h^{\alpha\beta}\!/h_{\alpha\beta}$. Integrating out the auxiliary  potential and further  fixing  the   coordinate gauge symmetry by ${\tx_{\mu}\equiv0}$~(\textit{c.f.~}\cite{Berman:2007vi}),  one    recovers      the familiar (Riemannian)  string action.

We now focus on   the maximally non-Riemannian  constant  background~(\ref{cHcJ}).  For simplicity, we ignore the $B$-field and diagonalize    the square matrices, $(X_{\mu}^{i},\brX_{\mu}^{\bri})$, $(Y^{\nu}_{i},\brY^{\nu}_{\bri})$, to  be  identity  matrices. The $D$-dimensional index, $\mu$, decomposes  into two parts: $\mu=(i,\bri)$. We perform a field redefinition of the potential, $A_{\alpha\mu}$, to a coordinate gauge symmetry invariant quantity, $p_{\alpha\mu}$,
\[
\ba{ll}
p_{\alpha i}:=\partial_{\alpha}\tx_{i}-A_{\alpha i}\,,\quad&\qquad
p_{\alpha \bri}:=A_{\alpha \bri}-\partial_{\alpha}\tx_{\bri}\,,
\ea
\] 
and prepare  a pair of   projection matrices from \cite{Green:1987sp}, 
\be
\ba{ll}
h_{\pm}^{\alpha\beta}=h_{\mp}^{\beta\alpha}=\half\!\left(h^{\alpha\beta}\pm{\frac{\epsilon^{\alpha\beta}}{\sqrt{-h}}}\right)\,,\quad&\quad\ba{l}
h_{\pm}{}_{\alpha}{}^{\beta}h_{\pm}{}_{\beta}{}^{\gamma}=h_{\pm}{}_{\alpha}{}^{\gamma}\,,\\
h_{\pm}{}_{\alpha}{}^{\beta}h_{\mp}{}_{\beta}{}^{\gamma}=0\,.
\ea
\ea
\label{OC}
\ee
The string Lagrangian~(\ref{stringaction})  now assumes   the form,
\be
\cL_{0\!}={-\sqrt{-h}}\!\left(
p_{\alpha i}h_{+}^{\alpha\beta}\partial_{\beta}x^{i}+
p_{\alpha \bri}h_{-}^{\alpha\beta}\partial_{\beta}x^{\bri}\right)
+\epsilon^{\alpha\beta}\partial_{\alpha}\tx_{\mu}\partial_{\beta}x^{\mu\,}.
\label{recast}
\ee
Evidently,  $p_{\alpha\mu}$'s are Lagrange multipliers  imposing  the chirality and anti-chirality  on the untilde coordinates: ${h_{+}^{\alpha\beta}\partial_{\beta}x^{i}\equiv0}$  and ${h_{-}^{\alpha\beta}\partial_{\beta}x^{\bri}\equiv0}$~\cite{Lee:2013hma,Morand:2017fnv} (\textit{c.f.~}\cite{Mason:2013sva,Siegel:2015axg,Lee:2017utr} which are fully chiral and equipped with a  Riemannian metric).

Toward the BRST quantization, it is convenient to   parametrize  ${\sqrt{-h}h^{\alpha\beta}}$, which has  unit determinant, by two  variables, without loss of generality, 
\[
\ba{lll}
\!{\sqrt{-h}h^{\tau\tau}}=-\frac{1}{e}\,,\quad&~~
{\sqrt{-h}h^{\tau\sigma}}=\frac{\omega}{e}\,,\quad&~~
{\sqrt{-h}h^{\sigma\sigma}}=e-\frac{\omega^{2}}{e}\,.
\ea
\]
Under diffeomorphisms, ${\delta_{c}\sigma^{\alpha}=c^{\alpha}}$,  these two  transform,
\be
\ba{l}
\delta_{c} e=c^{\alpha}\partial_{\alpha}e{+(}\partial_{\tau}c^{\tau}-\partial_{\sigma}c^{\sigma})e-2\partial_{\sigma}c^{\tau}\omega e\,,\\
\delta_{c}\omega\!=c^{\alpha}\partial_{\alpha}\omega{+(}\partial_{\tau}c^{\tau}-\partial_{\sigma}c^{\sigma})\omega+\partial_{\tau}c^{\sigma}-\partial_{\sigma}{c^{\tau}(}\omega^{2}+ e^{2})\,,
\ea
\label{deltaeo}
\ee
which  match  the standard  transformation of $\sqrt{-h}h^{\alpha\beta}$. We shall also  use (from time to time) the worldsheet light-cone convention,
\[
\ba{ll} 
\sigma^{\pm}=\tau\pm\sigma\,,\qquad&\qquad\partial_{\pm}=\half(\partial_{\tau}\pm\partial_{\sigma})\,,\\
c^{\pm}=c^{\tau}\pm c^{\sigma}\,,\qquad&\qquad p_{\pm\mu}=\half(p_{\tau\mu}\pm p_{\sigma\mu})\,.
\ea
\] 
In addition to the coordinate gauge symmetry~(\ref{CGS}) and the  worldsheet diffeomorphisms~(\ref{deltaeo}),  from (\ref{OC}), the Lagrangian~(\ref{recast})  admits extra gauge  symmetry: 
\be
\ba{lll}
\left\{\ba{l}
\delta p_{\alpha i}=h_{+}{}_{\alpha}{}^{\beta}\hC_{\beta i}\\
\delta p_{\alpha \bri}=h_{-}{}_{\alpha}{}^{\beta}\hC_{\beta \bri}
\ea
\right.\quad&~\Longleftrightarrow&\quad\left\{\ba{l}\delta p_{\pm i}=(\omega-e\pm 1)C_{i}\\
\delta p_{\pm \bri}=(\omega+e\pm 1)C_{\bri}
\ea\right.
\ea
\label{extra}
\ee
where $\hC_{\beta\mu}$'s are arbitrary local  parameters. Yet, despite their seemingly free index, \textit{i.e.~}`$\scriptstyle{\beta}$', since $h_{\pm\alpha}{}^{\beta}$'s are  ${2\times 2}$ projection matrices  with nontrivial kernel, the extra gauge symmetry can be specified  simply   by the alternative parameter, $C_{\mu}$, carrying no  worldsheet  index.

We proceed to fix all the  gauges, (\ref{CGS}), (\ref{deltaeo}), (\ref{extra}):
\be
\ba{lllll}
{e\equiv 1}\,,\quad&\quad {\omega\equiv0}\,,\quad&\quad {\tx_{\mu}\equiv0}\,,\quad&\quad
{p_{- i}\equiv0}\,,
\quad&\quad
{p_{+\bri}\equiv0}\,,
\ea
\label{GF}
\ee
which imply, ${-\sqrt{-h}h^{\alpha\beta}\equiv{\tiny{\left(\ba{cc}\mathbf{0}&\mathbf{1}\\\mathbf{1}&\mathbf{0}\ea\right)}}}$  on the light-cone and   the vanishing of  the topological term in (\ref{recast}).

The full Lagrangian with Faddeev--Popov ghosts is 
\be
\!\cL_{\scriptstyle{\bf{full\!}}}=\cL_{0}-i\deltaB\!\left(\ln e \,b_{e}+\omega b_{\omega}+\tx_{\mu}\tB^{\mu}+p_{-i}B^{i}+p_{+\bri}B^{\bri}\right)\,,
\label{FPL}
\ee
where $\{b_{e}, b_{\omega}, \tB^{\mu}, B^{\mu}\}$ are  the anti-ghosts   for  the gauge symmetries   of  (\ref{deltaeo}), (\ref{CGS}),  and (\ref{extra}).  With the associated    ghosts, $\{c^{\alpha}, \tC_{\mu}, C_{\mu}\}$,
and  auxiliary Nakanishi--Lautrup fields, $\{\kappa_{e},  \kappa_{\omega}, \tkappa^{\mu}, \kappa^{\mu}\}$, the BRST transformations are 
\be
\ba{c}
\deltaB x^{\mu}=c^{\alpha}\partial_{\alpha}x^{\mu}\,,\qquad\quad
\deltaB\tx_{\mu}=c^{\alpha}\partial_{\alpha}\tx_{\mu}+\tC_{\mu}\,,\\
\deltaB p_{\pm i}=(\omega{ -e}\pm 1)C_{i}+
c^{\alpha}\partial_{\alpha}p_{\pm i}+{\partial_{\pm}c^{+}}p_{+ i}
+{\partial_{\pm}c^{-}}p_{-i}\,,\\
\deltaB p_{\pm \bri}=(\omega {+e}\pm 1)C_{\bri}+
c^{\alpha}\partial_{\alpha}p_{\pm \bri}+{\partial_{\pm}c^{+}}p_{+ \bri}
+{\partial_{\pm}c^{-}}p_{-\bri}\,,\\
{}\deltaB c^{\alpha}=c^{\beta}\partial_{\beta}c^{\alpha}\,,\qquad\quad
{}\deltaB \tC_{\mu}=
c^{\alpha}\partial_{\alpha}\tC_{\mu}\,,\\
\deltaB C_{i}=
c^{\alpha}\partial_{\alpha}C_{i}+(\omega-e)\partial_{\sigma}c^{\tau}C_{i}+\partial_{\sigma}c^{\sigma}C_{i}\,,\\
\deltaB C_{\bri}=
c^{\alpha}\partial_{\alpha}C_{\bri}+(\omega+e)\partial_{\sigma}c^{\tau}C_{\bri}+\partial_{\sigma}c^{\sigma}C_{\bri}\,,\\
\deltaB b_{e}=i\kappa_{e}\,,~~\,
\deltaB b_{\omega}=i\kappa_{\omega}\,,~~\,
\deltaB \tB^{\mu}=i\tkappa^{\mu}\,,
~~\,\deltaB B^{\mu}=i{\kappa}^{\mu}\,,\\
\deltaB\kappa_{e}=
\deltaB\kappa_{\omega}=\deltaB\tkappa^{\mu}=
\deltaB{\kappa}^{\mu}=0\,,
\ea
\label{deltaBtxL}
\ee
while  $\deltaB e=\delta_{c}e$ and  $\deltaB\omega=\delta_{c}\omega$ are already given   in (\ref{deltaeo}), promoting the diffeomorphism parameters, $c^{\alpha}$, as ghosts. The transformations are off-shell nilpotent, ${\deltaB^{2}=0}$.

From the  variational  principle,   setting 
$b_{e}\equiv b_{++}{+b_{--}}$, $\,b_{\omega}\equiv b_{++}{-b_{--}}$,  and similarly for $\kappa_{e},\kappa_{\omega}$,  we acquire
\[
\ba{c}
p_{+i\,}\partial_{+}x^{i}+2ib_{++}\partial_{+}c^{+}+i(\partial_{+}b_{++})c^{+}=\kappa_{++}\,,\\
p_{-\bri\,}\partial_{-}x^{\bri}+2ib_{--}\partial_{-}c^{-}+i(\partial_{-}b_{--})c^{-}=\kappa_{--}\,,\\
p_{-i}=p_{+\bri}=\tkappa^{\mu}={\kappa}^{\mu}=\tB^{\mu}=\tC_{\mu}=B^{\mu}=C_{\mu}=0\,,
\ea
\label{vp1}
\]
and    the left/right-moving (chiral/anti-chiral) properties, 
\[
\ba{llll}
\partial_{-}x^{i}=0\,,\quad&\quad
\partial_{-}p_{+i}=0\,,\quad&\quad
\partial_{-}c^{+}=0\,,\quad&\quad
\partial_{-}b_{++}=0\,,\\
\partial_{+}x^{\bri}=0\,,\quad&\quad
\partial_{+}p_{-\bri}=0\,,\quad&\quad
\partial_{+}c^{-}=0\,,\quad&\quad
\partial_{+}b_{--}=0\,,
\ea
\label{vp2}
\]
which can be also derived from the reduced Lagrangian,
\be
\cL_{\scriptstyle{\bf{red.}}}= 2\big(p_{+i}\partial_{-}x^{i}+p_{-\bri}\partial_{+}x^{\bri}+ib_{++}\partial_{-}c^{+}+ib_{--}\partial_{+}c^{-}\big).
\ee
Naturally, $\{p_{+i}, p_{-\bri}\}$ are identified as the conjugate momenta of $\{x^{i},x^{\bri}\}$, forming  $D$ pairs of `$\beta\gamma$' system~\cite{Green:1987sp,Polchinski:1998rq} with the conformal weights  $1$ and $0$, for ${\beta_{i}\equiv p_{+i}}, {\brbeta_{\bri}\equiv p_{-\bri}}$ and ${\gamma^{j}\equiv x^{j}}, {\brgamma^{\brj}\equiv x^{\brj}}$ respectively.  Each pair  contributes to  a  central charge by two.   

The  BRST charge decomposes, $\QB=\QL+\QR$, with
\be
\ba{lll}
\!\QL&=\!&\displaystyle{\oint\rmd\sigma}~\beta_{i}\partial_{+}\gamma^{i}c^{+}+i(b_{++}\partial_{+}c^{+})c^{+}\\
{}&=\!&{\colon\sum_{m,n=-\infty}^{\infty}} n\left(-i\beta_{mi}\gamma^{i}_{n}+b_{m}c_{n}\right)c_{-m-n}\colon-ac_{0}\,,
\ea
\label{QB}
\ee
and  mirroring expression for $\QR$.  The quantization is given by 
 $
[\gamma_{m}^{i},\beta_{nj}]=i\delta^{i}_{~j}\delta_{m+n}$ and $\{b_{m},c_{n}\}=\delta_{m+n}$, which generate the normal ordering constant, $a$.   The BRST charges, $\QL,\QR$, are nilpotent,  if and only if ${n=\brn=13}$,  implying the usual  critical dimension, ${D=26}$,  since the  central charges are  ${\ccL=2n{-26}}$ and ${\ccR=2\brn{-26}}$, both of which should vanish. 

Physical states  are  annihilated by $\QL$ and the anti-ghost zero mode~$b_{0}$ (mirrored by  the right-moving sector).  Their  anti-commutator is
\be
L_{0}=\big\{b_{0},\QB\big\}=N_{\beta}{+N_{\gamma}}{+N_{b}}{+N_{c}}{-a}\,,
\label{acbQ}
\ee
where 
\[
\ba{ll}
N_{\beta}=\sum_{p=1}^{\infty}-ip\beta_{-pi}\gamma^{i}_{p}\,,\quad&\quad
N_{\gamma}={\sum_{p=1}^{\infty}}~ip\gamma_{-p}^{i}\beta_{pi}\,,\\
N_{b}={\sum_{p=1}^{\infty}}~pb_{-p}c_{p}\,,\quad&\quad
N_{c}={\sum_{p=1}^{\infty}}~pc_{-p}b_{p}\,,
\ea
\]
are the level-counting  operators for each   creation operator with 
${p\geq 1}$. These are all positive semi-definite. Hence, the vanishing  of $L_{0}$~(\ref{acbQ}) on physical states means    a drastic truncation of the entire   string spectrum to just one level.  Computing     $\langle0|[L_{1},L_{-1}]|0\rangle=-2$ with $L_{n}=\{\QL,b_{n}\}$, we identify the level to be unity, ${a=1}$. Then, from $\beta_{0i}|k{\rangle=\,}k_{i}|k\rangle$, $\QL c_{-1}|k{\rangle=}\,\QL\gamma_{-1}^{i}|k{\rangle=0}$,  and
\[
\ba{ll}
\!\QL\beta_{-1i}|k\rangle=k_{i}c_{-1}|k\rangle\,,\quad&
\QL b_{-1}|k\rangle=(ik_{i}\gamma^{i}_{-1}+2c_{-1}b_{0})|k\rangle\,,
\ea
\]
the physical states consist of four sectors, 
(with  $|{k\!\downarrow\rangle}$ satisfying $b_{0}|{k\!\downarrow}{\rangle=0}$~\cite{Green:1987sp,Polchinski:1998rq}),
\[
\ba{ll}
\delta\cH_{i\bri}\,
\gamma_{-1}^{i}|{k_{j}\!\downarrow}{\rangle\otimes} \brgamma_{-1}^{\bri}|{k_{\brj}\!
\downarrow}\rangle\,,\quad&~~
\delta\cH_{i}{}^{\bri}\,
\gamma_{-1}^{i}|{k_{j}\!\downarrow}{\rangle\otimes} \brbeta_{-1\bri}|{k_{\brj}\!
\downarrow}\rangle\,,\\
\delta\cH^{i}{}_{\bri}\,
\beta_{-1i}|{k_{j}\!\downarrow}{\rangle\otimes} \brgamma_{-1}^{\bri}|{k_{\brj}\!
\downarrow}\rangle\,,\quad&~~
\delta\cH^{i\bri}\,
\beta_{-1i}|{k_{j}\!\downarrow}{\rangle\otimes}\brbeta_{-1\bri}|{k_{\brj}\!
\downarrow}\rangle\,,
\ea
\]
which should satisfy on-shell relations for  $\QB$-closedness,
\be
\ba{llll}
k_{\bri}\delta\cH_{i}{}^{\bri}=0\,,\quad&~~
k_{i}\delta\cH^{i}{}_{\bri}=0\,,\quad&~~
k_{i}\delta\cH^{i\bri}=0\,,\quad&~~
k_{\bri}\delta\cH^{i\bri}=0\,,
\ea
\label{QOS}
\ee
and equivalence relations as for gauge symmetries,
\be
\ba{cc}
\multicolumn{2}{c}{\delta\cH_{i\bri}\sim\delta\cH_{i\bri}+
k_{i}\lambda_{\bri}- k_{\bri}\lambda_{i}\,,}\\
 \delta\cH_{i}{}^{\bri}\sim \delta\cH_{i}{}^{\bri}+k_{i}\xi^{\bri}\,,\qquad&\qquad
  \delta\cH^{i}{}_{\bri}\sim \delta\cH^{i}{}_{\bri}- k_{\bri}\xi^{i}\,,
\ea
\label{QER}
\ee 
where $\xi^{i},\xi^{\bri}$ need to be  divergenceless,    $k_{i}\xi^{i}=k_{\bri}\xi^{\bri}=0$.

We have  a good reason to denote the physical states by the same symbol as the generalized metric: the $4n\brn$ of  $\left\{\delta\cH_{i\bri},
\delta\cH_{i}{}^{\bri},\delta\cH^{i}{}_{\bri},\delta\cH^{i\bri}\right\}$ are literally   the moduli of the  generalized metric~(\ref{cHcJ}) that we have been dealing with,  in the  diagonal form where the only nontrivial components are $\cH_{i}{}^{j}=\cH^{j}{}_{i}=\delta_{i}{}^{j}$ and $\cH_{\bri}{}^{\brj}=\cH^{\brj}{}_{\bri}=-\delta_{\bri}{}^{\brj}$~\cite{Cho:2019ofr}. On-shell, they meet the linearized DFT equations of motion~\cite{Ko:2015rha} (see also \cite{Hohm:2015ugy}):
\be
\ba{ccc}
\partial_{i}\partial_{j}\delta\cH^{j\bri}=0\,,\quad&\quad
\partial_{\bri}\partial_{\brj}\delta\cH^{i\brj}=0\,,\quad&\quad
\partial_{i}\partial_{\bri}\delta\cH^{i\bri}=0\,,\\
\multicolumn{3}{c}{
\partial_{i}\partial_{j}\delta\cH^{j}{}_{\bri}-\partial_{\bri}\partial_{\brj}\delta\cH_{i}{}^{\brj}+4\partial_{i}\partial_{\bri}\delta d=0\,,}
\label{lEDFE}
\ea
\ee
which enjoy local symmetries   inherited from the General Covariance of DFT (generalized Lie derivative, $\hcL_{\xi}\cH_{MN}$),
\be
\ba{ll}
\delta(\delta\cH^{i}{}_{\bri})=\partial_{\bri}\xi^{i}\,,\quad&\quad
\delta(\delta\cH_{i}{}^{\bri})=-\partial_{i}\xi^{\bri}\,,\\
\delta(\delta\cH_{i\bri})=\partial_{\bri}\lambda_{i}-\partial_{i}\lambda_{\bri}\,,
\quad&\quad
\delta(\delta d)=-\quarter(\partial_{i}\xi^{i}+\partial_{\bri}\xi^{\bri})\,.
\ea
\label{ldiffeo}
\ee
We may  choose a gauge, ${\delta d=0}$. Remarkably then, (\ref{QOS}), (\ref{QER}) imply (\ref{lEDFE}), (\ref{ldiffeo}). Further, restricted to   normalizable solutions, the converse appears also true. The first and second   in (\ref{lEDFE}) give $\partial_{j}\delta\cH^{j\bri}=0,\, \partial_{\brj}\delta\cH^{i\brj}=0$, which are   generically solved by $\delta\cH^{i\bri}=\epsilon^{ijkl\cdots}\epsilon^{\bri\brj\brk\brl\cdots}\partial_{j}\partial_{\brj}\Phi_{kl\cdots\brk\brl\cdots}$, and hence the third holds. The last implies $\partial_{i}\partial_{j}\delta\cH^{j}{}_{\bri}=\partial_{\bri}\partial_{\brj}\delta\cH_{i}{}^{\brj}=\partial_{i}\partial_{\bri}\varphi$ for some $\varphi$. Again for normalizable solutions, we  get $\partial_{j}\delta\cH^{j}{}_{\bri}=\partial_{\bri}\varphi$ and $\partial_{\brj}\delta\cH_{i}{}^{\brj}=\partial_{i}\varphi$, which can be gauged away using the remaining  (\ref{ldiffeo}).  

As the spectrum is finite,  DFT itself is to be identified as string field theory around the non-Riemannian vacua. Evident from the position of the indices,  it is $\delta\cH^{i\bri}$ that may condensate and reduce the `non-Riemannianity', to build up Riemannian spacetime~\cite{Cho:2019ofr}.

\section*{Comments: supersymmetric extension}
\noindent Our BRST charge formula~(\ref{QB}) can be  easily  extended to     a generic $(n,\brn)$ non-Riemannian background, to include $n$ pairs of chiral $\beta\gamma$, $\brn$ pairs of anti-chiral  ${\brbeta}{\brgamma}$, and   ordinary  (left-right combined) $D{-n}{-\brn}$ number of  $x^{\mu}$.  The central charges are  $\ccLR=D\pm(n-\brn)-26$,   and  thus  necessarily  ${n=\brn}$ and ${D=26}$. Non-relativistic string theories~\cite{Gomis:2000bd,Danielsson:2000gi,Andringa:2012uz,Harmark:2017rpg} are    examples  of     ${(n,\brn)=(1,1)}$~\cite{Kim:2007hb}.   

The necessity of putting ${n=\brn}$ was also noted in \cite{Morand:2017fnv} as a  condition to embed  non-Riemannian geometries into type II doubled superstring	\cite{Park:2016sbw,Blair:2019qwi} or supersymmetric  DFTs~\cite{Hohm:2011zr,Jeon:2011sq,Jeon:2012hp}, the  constructions of which rely  on  genuine $\ODD$ covariant vielbeins rather than the Riemannian   zehnbein, $e_{\mu}{}^{a}$~\cite{Coimbra:2011nw}.  The central charges  should  be   $\ccLR=D\pm(n-\brn)-10$,   indicating   that  ${n=\brn}$ non-Riemannian geometries are consistent superstring vacua  in ${D=10}$, which enlarges  the string theory landscape far  beyond the Riemannian paradigm.

Chiral string means $x^{i}(\tau,\sigma)=x^{i}(0,\tau+\sigma)$: it is fixed in space and  thus hardly interacts  with one another. This classical intuition may suggest    us   to   explore  non-Riemannian backgrounds  (either  flat or curved)  as candidates for an  internal space,  alternative to string compactification traditionally on ``small" Riemannian manifolds.     At a glance, in the presence of external four-dimensional  Minkowskian spacetime, the  truncation  of the string spectrum to just one level may be no longer the case as $L_{0}$ will include external $p_{\mu}p_{\nu}\eta^{\mu\nu}$ which is not positive definite. Nevertheless,  since $[x^{i}(\tau_{1},\sigma_{1}),x^{j}(\tau_{2},\sigma_{2})]=0$, 
the  correlation functions of bosonic string tachyon vertex operators will have trivial   dependency along all non-Riemannian directions: with $k\cdot x=k_{\mu}x^{\mu}+k_{j}x^{j}+k_{\brj}x^{\brj}$, 
\[
\bigg\langle\prod_{a=1}^{N}\colon e^{ik_{a}\cdot\, x(\tau_{a},\sigma_{a})}\colon
 \bigg\rangle=\delta^{D\!}\!\left(
\textstyle{\sum_{b=1}^{N}}k_{b}\right)F(k_{a\mu},\tau_{a},\sigma_{a})\,,
 \]
where $F(k_{a\mu},\tau_{a},\sigma_{a})$ is independent of  $k_{aj},k_{a\brj}$'s. This  indicates   delta-function interactions   on the internal  $x$-space after  Fourier transformation,  and thus  is consistent with   the classical intuition of the chiral string.\\

\noindent\textit{Acknowledgments.}  {We  are deeply grateful to  Taichiro Kugo for encouragement.  JHP also thanks  Chris  Blair and  Gerben Oling  for discussion. The work of JHP  was  supported by Basic Science Research Program by the National Research Foundation of Korea (NRF)   through  the Grants,  NRF-2016R1D1A1B01015196 and NRF-2020R1A6A1A03047877. The work of SS was supported by JSPS KAKENHI (Grant-in-Aid for Scientific Research (B)) grant number JP19H01897.}
\hfill

\appendix
\setlength{\jot}{9pt}                 
\renewcommand{\arraystretch}{2} 

\end{document}